\documentclass[preprint]{sig-alternate-05-2015}
\usepackage{amsmath}
\usepackage{amssymb}
\usepackage{cmll}
\usepackage{dsfont}
\usepackage{url}
\usepackage{enumitem}
\usepackage{listings}
\usepackage{algorithm}
\usepackage{algorithmic}
\usepackage{endnotes}
\usepackage{graphicx}

\usepackage{booktabs}
\usepackage{multirow}
\usepackage{siunitx}

\usepackage[font=scriptsize,labelfont=bf]{caption}

\graphicspath{ {images/} }

\conferenceinfo{PaPoC'16}{April 18, 2016, London, UK}
\CopyrightYear{2016}
\global\boilerplate={Copyright is held by the owner/author(s). Publication rights licensed to ACM.}
\crdata{978-1-4503-2716-9/14/04}

\title{Big(ger) Sets: decomposed delta CRDT Sets in Riak}
\subtitle{[Work in progress report]}

\numberofauthors{3}
\author{
\alignauthor
Russell Brown\\
       \affaddr{Basho Technologies, Ltd}\\
       \affaddr{United Kingdom}\\
       \email{russelldb@basho.com}
\alignauthor
Zeeshan Lakhani\\
       \affaddr{Basho Technologies, Inc}\\
       \affaddr{USA}\\
       \email{zlakhani@basho.com}
\alignauthor
Paul Place\\
       \affaddr{Basho Technologies, Inc}\\
       \affaddr{USA}\\
       \email{pplace@basho.com}
}

\let\footnote=\endnote

\begin{document}
\CopyrightYear{2016}
\setcopyright{acmlicensed}
\conferenceinfo{PaPOC'16,}{April 18-21 2016, London, United Kingdom}
\isbn{978-1-4503-4296-4/16/04}\acmPrice{\$15.00}
\doi{http://dx.doi.org/10.1145/2911151.2911156}
\maketitle

\begin{figure*}[t]
\centering
\includegraphics[width=14cm, height=8cm]{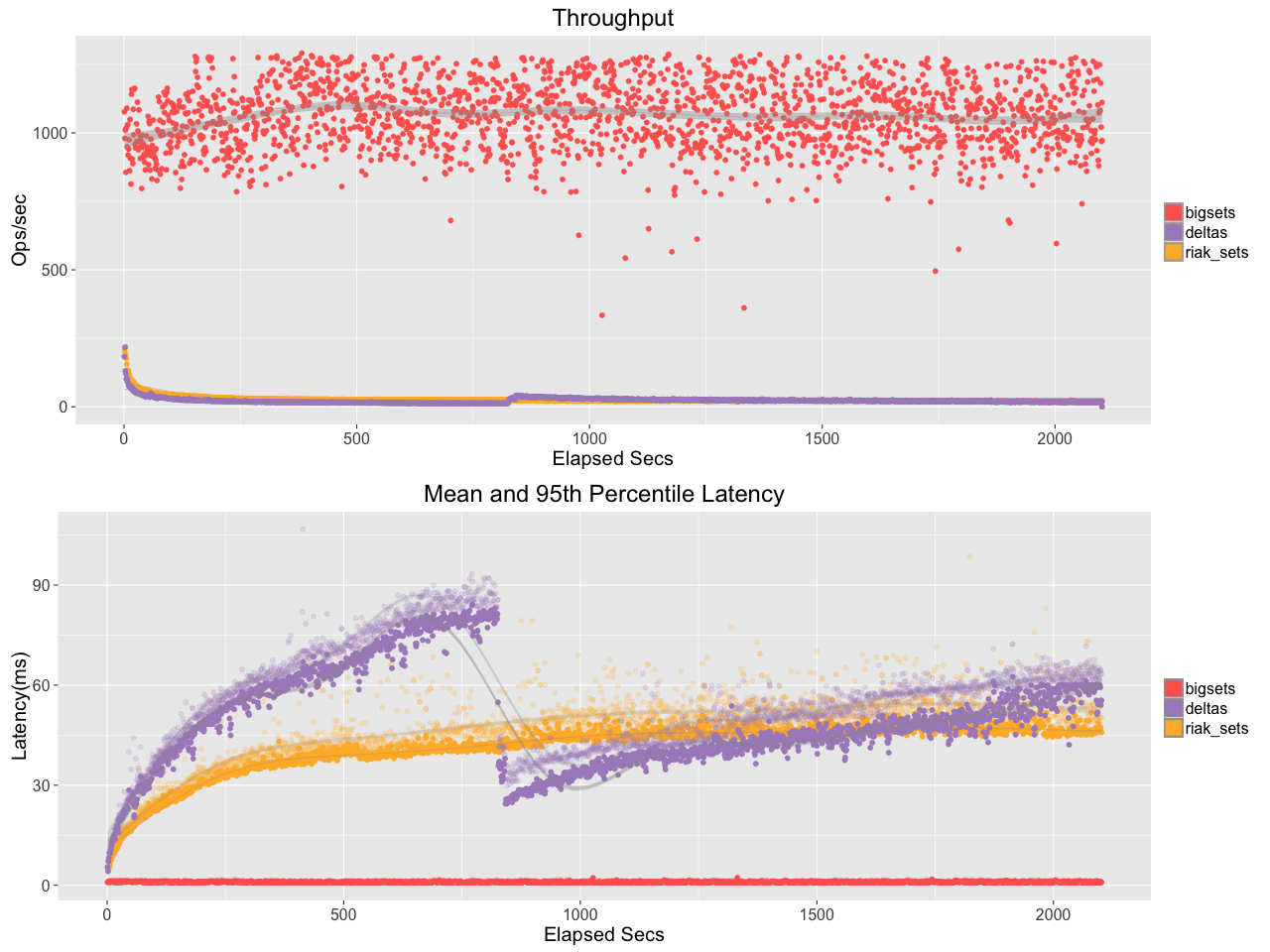}
\caption{Comparative benchmark using Basho Bench\cite{basho_bench}, inserting
  up-to 10,000 4-byte elements for 35 minutes.}
\label{figuremain}
\end{figure*}

\begin{abstract}
CRDT\cite{DBLP:conf/sss/ShapiroPBZ11} Sets as implemented in
Riak\cite{riak} perform poorly for writes, both as cardinality grows,
and for sets larger than 500KB\cite{kms}. Riak users wish to create
high cardinality CRDT sets, and expect better than $O(n)$ performance
for individual insert and remove operations. By decomposing a CRDT set
on disk, and employing
delta-replication\cite{DBLP:conf/netys/AlmeidaSB15}, we can achieve
far better performance than just delta replication alone: relative to
the size of causal metadata, not the cardinality of the set, and we
can support sets that are 100s times the size of Riak sets, while
still providing the same level of consistency. There is a trade-off in
read performance but we expect it is mitigated by enabling queries on
sets.
\end{abstract}

%\category{H.4}{Information Systems Applications}{Miscellaneous}

\section{Introduction}

Riak is an eventually consistent\cite{Vogels:2009:EC:1435417.1435432}
key-value database inspired by
Dynamo\cite{DeCandia:2007:DAH:1323293.1294281}. To ensure write
availability, Riak allows concurrent writes to keys. Some users find
resolving conflicts resulting from concurrency hard\cite{bet365} so we
added some CRDTs\cite{shapiro:inria-00555588} to Riak. Our initial
implementation was to create an open source library of purely functional
datatypes\cite{riakdt} and embed them in Riak with an
API\cite{riakdatatypes}. In this paper we consider only the Set, an
Erlang implementation of the state-based
ORSWOT\cite{DBLP:journals/corr/abs-1210-3368} (Observe-Remove-Set-Without-
Tombstones).

In this paper we show that:

\begin{enumerate}[noitemsep, nolistsep]
    \item a naive implementation of data types in Riak performs
      poorly, especially for larger cardinality sets (section \ref{motivate})
    \item simply adding delta replication is not much better (section
      \ref{delta})
    \item decomposing a set CRDT into its constituent parts (logical
      clock and set members) yields great improvements in write
      performance (section \ref{bigset})
\end{enumerate}

This final point above is our work's primary contribution, and speaks
to the needs of practitioners to do more than translate research into
code, and consider the lifetime of the CRDT in the system, including
such mundane details as durable storage. See figure \ref{figuremain} if
you read nothing else. While decomposing the set may itself seem a
reasonably obvious approach, it raises issues around anti-entropy and
reads that require creative solutions.

\section{Motivation}\label{motivate}

When Basho released Riak DataTypes in Riak 2.0\cite{riak2.0}, the
first thing users did was treat Riak Sets like Redis Sets\cite{redis}
and try and store millions of elements in a set. Redis is a
non-distributed, in-memory data structure server, not a distributed,
fault-tolerant key/value database. Riak was unable to perform
satisfactorily to user expectations when performing inserts into sets,
especially as set cardinality grew.

Riak stores all data in riak-objects, a kind of
multi-value-register\cite{shapiro:inria-00555588}. As Riak stores each
object durably on disk in a local key/value store (either bitcask
or leveldb) there is a limit to the size each key/value pair can
be\cite{romax}, and since Riak's CRDTs are stored in a riak-object,
they inherit that limit.

A riak-object has a version vector\cite{Fidge:1991:LTD:112827.112860},
and an opaque binary payload. The CRDT is stored in the binary
payload.

Users report a degradation in write performance as sets grow\cite{kms}.
In all the literature a CRDT is a thing in memory. There is only one
of them, and every actor is a replica. In a client/server database this is
not the case: a database services many clients, and stores many objects
durably on disk. CRDTs in Riak must therefore be serialized for both storage
and transfer over the network.

\subsection{Understanding Riak Set Performance}
When a client wishes to add an element to a Riak set it sends an
operation: '\lstinline!add element E to set S!'

Riak forwards the operation to a coordinating replica that will: read
a riak-object from disk, de-serialize the set, add the element to the
set, serialize the set, update the riak-object's version vector, write
the riak-object to disk, and send the riak-object downstream for
replication.

Each replica will then read their local copy from disk. If the
incoming update supersedes their local copy, determined by comparing
riak-object version vectors, they simply store the incoming
riak-object. If conflict is detected, the set is de-serialized from the
riak-object, the CRDT join function is called, and the result
serialised, the riak-object version vectors are merged, and the
riak-object is finally written to disk.

The key insight is that the set is stored in a single object. To add
an element, Riak must read and write the whole object. This dictates
both performance and the size to which a set can grow. Per-operation
this yields an $O(n)$ cost, where $n$ is the size of the set. However,
over the life of the set the cost is $O(n^2)$, both in terms of bytes
read and written, and transfered over the network. Where a single
insert is $O(n)$ filling a set by single inserts from $0\cdots$$n$ is
$O(n^2)$!

\section{Delta CRDTs}\label{delta}

State based CRDTs are join-semi-lattices. They work by ``joining'' or
``merging'' divergent replica state into a single value in a
deterministic way. As described above in section \ref{motivate}, Riak
uses ``full state replication'' to propagate updates to
Sets. Delta-CRDTs\cite{DBLP:conf/netys/AlmeidaSB15} are a CRDT variant
that promise a more effecient network utilisation for replication of
updates. In order to improve performance we implemented
delta-datatypes\cite{riakdtdelta}. A delta-CRDT is one that, when
updated, generates a smaller-than-full-state delta-state that can be
replicated. The performance in Riak was virtually unchanged. As per
the paper the upstream replica generates a delta and sends it
downstream. Downstream the set \textbf{\textit{must always}} merge the
delta as an incoming delta \textbf{\textit{never}} supersedes the
local state, even without concurrency! The saving in terms of network
transmission comes at the expense of always deserialising and merging
downstream.

\section{BigSet: An overview}\label{bigset}

We wrote a prototype system, \textbf{\textit{bigset}}, to address
these issues.

In bigset, as in Riak, the actors in the system are called vnodes\cite{vnodes}.
\textit{N} vnodes each store a replica of each datum. Vnodes may
act concurrently. Vnodes service requests for many clients; vnodes
store many data items.

The key innovation is that bigset decomposes the Set CRDT across a
range of keys. An ORSWOT CRDT Set is composed of a logical clock,
opaque set members, and per-member causal information called
``dots''. Rather than mapping a key (a set name) to an object that
contains the whole CRDT Set, Bigset gives each of these component
elements of a CRDT Set their own key in an ordered durable key/value
datastore.

A bigset is made of a set-clock, a set-tombstone, and a collection of
element-keys.

\subsection{Clocks}

Both the set-clock and set-tombstone are logical clocks. The former
summarises events seen by the actor, the latter events removed by the
actor. The set-clock grows as events occur and are replicated in the
system. The set-tombstone temporalily grows as it records removal
information, and then shrinks as keys are discarded.

The clocks consist of a 2-tuple of
\lstinline[language=erlang]!{BaseVV(), DotCloud()}!  Where
\lstinline!BaseVV! is a normal Version Vector and \lstinline!DotCloud!
is a mapping of ActorIDs to a list of integers which denote the events
seen at the replica that are not contiguous to the base Version
Vector\cite{DBLP:conf/netys/AlmeidaSB15}\cite{DBLP:journals/dc/MalkhiT07}. A
replica will \textbf{never} have an entry for itself in the DotCloud.

\subsection{Elements}

Elements are the opaque values that clients wish to store in a Set. In
bigset each insert of an element gets its own key in
leveldb\cite{leveldb}. The key is a composite made of the set name,
the element, and a dot\cite{1466}. The dot is a pair of $(actorId,
Event)$ denoting the logical event of inserting the element. The
element-keys are also the delta that is replicated.

\subsection{Writes}

When writing, bigset does not read the whole set from disk; instead, it
reads the set's logical clocks only and writes the updated clock(s)
and any element-keys that are being inserted.  For removes we need
only read and write the clocks. This is how bigset acheives its write
performance.

\textbf{NOTE}: multiple elements can be added/removed at once, but for
brevity/simplicity we discuss the case of single element adds/removes
only.

\subsubsection{Inserts}

When a client wishes to insert an element it sends an operation:
'\lstinline!add element E to set S with Ctx!'. It may provide a causal
context (hereafter just context) for the operation, which provides
information as to what causal history makes up the Set the client has
observed. The context will be empty if the client has not seen element
$E$ at all (as we expect to be the common case.)\footnote{Why a
  context?  Adds are removes. If the client has already seen some
  element $E$ in the set a new insert of $E$ replaces the old.  Adding
  element $E$ at replica $X$ will cause all dots for $E$ at $X$ to be
  removed and the single new event will be the sole surviving dot for
  $E$. This optimisation comes from the reference
  implementation\cite{cbdelta} and assumes that the actor
  \textbf{\textit{is}} a replica. With action-at-a-distance it is more
  complex: the client is not a replica, so must tell the coordinating
  replica what it has seen, it must read the set (or at least the
  element) to get a context. To clarify why, imagine a client reads
  from vnode $A$, but writes to vnode $B$.}

Bigset sends this operation to a coordinating vnode that will run
algorithm \ref{alg1} (see \ref{appendix:algorithms}). Briefly: It
reads the clock for the set, increments it, creates a key for the new
element, and stores it and the clock, atomically.

It then sends the new key downstream as a ``delta''. The downstream
replicas run algorithm \ref{alg2} (see \ref{appendix:algorithms}):
read the local clock, if the local clock has seen the delta's dot,
then it is a no-op, otherwise the clock adds the new dot to itself,
and stores the updated clock and delta atomically.

\subsubsection{Removes}

Removes are as per-writes except there is no need to coordinate, and
no element key to write. To have an effect the client
\textbf{\textit{must}} provide a context for a remove. The remove
algorithm is far simpler: if the set-clock has seen the context, add
the dots of the context to the set-tombstone. Otherwise, add them to
the set-clock. This ensures that, if the adds were unseen they never
get added, and if they were seen, they will get compacted out (see
\ref{compaction} below).

\subsubsection{Compaction} \label{compaction}

We use leveldb\cite{leveldb-basho} to store bigsets. Leveldb has a
mechanism to remove deleted keys called compaction\cite{compact}. We
have modified leveldb to use the set-tombstone in compaction. As
leveldb considers a key $K$ for compaction it uses the
set-tombstone. If $K.dot$ is seen by the tombstone, the key is
discarded. This way we remove superseded/deleted keys without issuing
a delete. Once a key is removed the set-tombstone subtracts the
deleted dot. This trims the set-tombstone, keeping its size minimal.

\subsection{Reads} \label{reads}

A bigset read is a leveldb iteration, or fold, over the keyspace for a
set. It transforms the decomposed bigset into a traditional ORSWOT
state-based CRDT. Every element key that is NOT covered by the
set-tombstone is added to the ORSWOT.

As bigsets can be {\textit{big}}, we don't wait to build the entire
set before sending to the client, we stream the set in configurable
batches (by default $10000$ elements at a time.)

Riak, and bigsets, allow clients to read from a quorum of
replicas. Bigset has a novel streaming ORSWOT CRDT Join operation,
that is able to perform a merge on subsets of an ORSWOT. This is
enabled by the fact that the set element keys are stored and therfore
streamed in lexicographical element order. This ordering and
incremental merging allows us to query a bigset. We can discover if
$X$ is a member of Set without reading the whole set. We can seek to a
point in the set and stream, to enable pagination or range queries. We
expect this to mitigate the current negative performance delta between
Riak DataTypes and bigset for reads.

\section{Experience With Bigsets}

Not yet in production, but being developed, we do have both initial
benchmark results, and property based tests for bigsets. Our property
based tests using quickcheck\cite{eqc} show that bigset and Riak sets
are semantically equivalent. The benchmarks show us that bigsets
performs far better for writes (see figure \ref{figuremain}), while
paying a read performance penalty (see figure \ref{figure1kreads}) which we plan
to engineer our way out of with low-level code, and by providing queries
over the Set, making full set reads unnecessary in most cases.

\section{Summary}

We've characterised the key difference between bigsets and Riak
DataType sets: decomposition of the CRDTs structure, minimal
read-before-write, and a set-tombstone and set-clock that make joining
a delta as simple as adding its causal data to a clock and appending
the key to leveldb.

The poor performance of CRDTs in Riak led to the bigsets design, which
clearly demonstrates that considering the primary need to durably
store a CRDT means optimising for bytes read and written. We have much
work to do to bring this prototype into the Riak KV database. We plan
in the future to write more about key processes we have developed
including anti-entropy and hand-off, and also generality for
application to other data types, including Riak
Maps\cite{Brown:2014:RDM:2596631.2596633}.

\section{Acknowledgements}
Many thanks to Scott Lystig Fritchie and John Daily for their time
reviewing drafts.

Funding from the European Union Seventh Framework Program
(FP7/2007-2013) under grant agreement 609551, SyncFree project.

\bibliographystyle{abbrv}
\bibliography{sigproc}

\begin{thebibliography}{10}

\bibitem{1466}
P.~S. Almeida, C.~B. Moreno, R.~Gon{\c c}alves, N.~Pregui{\c c}a, and V.~Fonte.
\newblock Scalable and accurate causality tracking for eventually consistent
  stores.
\newblock In {\em Distributed Applications and Interoperable Systems}, volume
  8460, Berlin, Germany, June 2014. Springer, Springer.

\bibitem{DBLP:conf/netys/AlmeidaSB15}
P.~S. Almeida, A.~Shoker, and C.~Baquero.
\newblock Efficient state-based crdts by delta-mutation.
\newblock In A.~Bouajjani and H.~Fauconnier, editors, {\em Networked Systems -
  Third International Conference, {NETYS} 2015, Agadir, Morocco, May 13-15,
  2015, Revised Selected Papers}, volume 9466 of {\em Lecture Notes in Computer
  Science}, pages 62--76. Springer, 2015.

\bibitem{cbdelta}
C.~Baquero.
\newblock Delta crdts.
\newblock \url{https://github.com/CBaquero/delta-enabled-crdts}.

\bibitem{basho_bench}
Basho.
\newblock Basho bench.
\newblock \url{http://docs.basho.com/riak/latest/ops/building/benchmarking/}.

\bibitem{leveldb-basho}
Basho.
\newblock Leveldb.
\newblock
  \url{http://docs.basho.com/riak/latest/ops/advanced/backends/leveldb/}.

\bibitem{riak}
Basho.
\newblock Riak.
\newblock \url{http://basho.com/products/riak-kv/}.

\bibitem{riak2.0}
Basho.
\newblock Riak 2.0.
\newblock \url{http://basho.com/posts/technical/introducing-riak-2-0/}.

\bibitem{riakdatatypes}
Basho.
\newblock Riak datatypes.
\newblock \url{https://docs.basho.com/riak/2.1.3/dev/using/data-types/}.

\bibitem{riakdt}
Basho.
\newblock Riak dt.
\newblock \url{https://github.com/basho/riak_dt}.

\bibitem{riakdtdelta}
Basho.
\newblock Riak dt deltas.
\newblock \url{https://github.com/basho/riak_dt/tree/delta_data_types}.

\bibitem{romax}
Basho.
\newblock Riak object max size.
\newblock
  \url{http://docs.basho.com/riak/latest/community/faqs/developing/#is-there-a-limit-on-the-size-of-files-that-can-be}.

\bibitem{vnodes}
Basho.
\newblock Vnodes.
\newblock \url{http://docs.basho.com/riak/latest/theory/concepts/vnodes/}.

\bibitem{DBLP:journals/corr/abs-1210-3368}
A.~Bieniusa, M.~Zawirski, N.~M. Pregui{\c{c}}a, M.~Shapiro, C.~Baquero,
  V.~Balegas, and S.~Duarte.
\newblock An optimized conflict-free replicated set.
\newblock {\em CoRR}, abs/1210.3368, 2012.

\bibitem{Brown:2014:RDM:2596631.2596633}
R.~Brown, S.~Cribbs, C.~Meiklejohn, and S.~Elliott.
\newblock Riak dt map: A composable, convergent replicated dictionary.
\newblock In {\em Proceedings of the First Workshop on Principles and Practice
  of Eventual Consistency}, PaPEC '14, pages 1:1--1:1, New York, NY, USA, 2014.
  ACM.

\bibitem{leveldb}
J.~Dean and S.~Ghemawat.
\newblock Leveldb.
\newblock \url{https://rawgit.com/google/leveldb/master/doc/index.html}.

\bibitem{DeCandia:2007:DAH:1323293.1294281}
G.~DeCandia, D.~Hastorun, M.~Jampani, G.~Kakulapati, A.~Lakshman, A.~Pilchin,
  S.~Sivasubramanian, P.~Vosshall, and W.~Vogels.
\newblock Dynamo: Amazon's highly available key-value store.
\newblock {\em SIGOPS Oper. Syst. Rev.}, 41(6):205--220, Oct. 2007.

\bibitem{Fidge:1991:LTD:112827.112860}
C.~Fidge.
\newblock Logical time in distributed computing systems.
\newblock {\em Computer}, 24(8):28--33, Aug. 1991.

\bibitem{compact}
Google.
\newblock Leveldb file layout and compactions.
\newblock \url{https://leveldb.googlecode.com/svn/trunk/doc/impl.html}.

\bibitem{bet365}
D.~Macklin.
\newblock Key lessons learned from transition to nosql at an online gambling
  website.
\newblock
  \url{http://www.infoq.com/articles/key-lessons-learned-from-transition-to-nosql}.

\bibitem{DBLP:journals/dc/MalkhiT07}
D.~Malkhi and D.~B. Terry.
\newblock Concise version vectors in winfs.
\newblock {\em Distributed Computing}, 20(3):209--219, 2007.

\bibitem{eqc}
T.~A. QuviQ, John~Hughes.
\newblock Erlang-quickcheck.
\newblock \url{http://www.quviq.com/products/erlang-quickcheck/}.

\bibitem{redis}
S.~Sanfilippo.
\newblock Redis.
\newblock \url{https://en.wikipedia.org/wiki/Redis}.

\bibitem{shapiro:inria-00555588}
M.~Shapiro, N.~Pregui{\c c}a, C.~Baquero, and M.~Zawirski.
\newblock {A comprehensive study of Convergent and Commutative Replicated Data
  Types}.
\newblock Research Report RR-7506, {Inria -- Centre Paris-Rocquencourt ;
  INRIA}, Jan. 2011.

\bibitem{DBLP:conf/sss/ShapiroPBZ11}
M.~Shapiro, N.~M. Pregui{\c{c}}a, C.~Baquero, and M.~Zawirski.
\newblock Conflict-free replicated data types.
\newblock In X.~D{\'{e}}fago, F.~Petit, and V.~Villain, editors, {\em
  Stabilization, Safety, and Security of Distributed Systems - 13th
  International Symposium, {SSS} 2011, Grenoble, France, October 10-12, 2011.
  Proceedings}, volume 6976 of {\em Lecture Notes in Computer Science}, pages
  386--400. Springer, 2011.

\bibitem{kms}
K.~M. Spartz.
\newblock Benchmarking large riak data types.
\newblock
  \url{http://kyle.marek-spartz.org/posts/2014-12-02-benchmarking-large-riak-data-types-continued.html}.

\bibitem{Vogels:2009:EC:1435417.1435432}
W.~Vogels.
\newblock Eventually consistent.
\newblock {\em Commun. ACM}, 52(1):40--44, Jan. 2009.

\bibitem{fork}
Basho.
\newblock Basho OTP.
\newblock
  \url{https://github.com/basho/otp/releases/tag/OTP_R16B02_basho8}.

\bibitem{pareto}
J.~L. Petersen.
\newblock Estimating the Parameters of a Pareto Distribution.
\newblock
  \url{http://www.math.umt.edu/gideon/pareto.pdf}.

\bibitem{ec2}
Amazon.
\newblock Previous Generation Instances.
\newblock
  \url{http://aws.amazon.com/ec2/previous-generation}.

\end{thebibliography}

\theendnotes

\section{Appendix: algorithms} \label{appendix:algorithms}
\begin{algorithm}
\caption{bigset coordinate insert}
\label{alg1}
\begin{algorithmic}
  \REQUIRE set S, element E, op-context Ctx
  \STATE SC = read set-clock
  \STATE TS = read set-tombstone
	\FOR{Dot in Ctx}
	\IF{SC  not seen Dot}
		\STATE SC = join(SC, Dot)
	\ELSE
		\STATE TS = join(TS, Dot)
		\ENDIF
	\ENDFOR
    \STATE SC.increment()
    \STATE Dot = SC.latest-dot()
    \STATE Val = (S, E, Dot)
    \STATE atomic-write([SC, TS, Val])
    \STATE send-downstream(Val, Ctx)
\end{algorithmic}
\end{algorithm}

\begin{algorithm}
\caption{bigset replica insert}
\label{alg2}
\begin{algorithmic}
  \REQUIRE set S, val V, op-context Ctx
  \STATE SC = read set-clock
  \STATE TS = read set-tombstone
	\FOR{Dot in Ctx}
	\IF{SC  not seen Dot}
		\STATE SC.join( Dot)
	\ELSE
		\STATE TS .join(Dot)
		\ENDIF
	\ENDFOR
	\IF{SC not seen V.dot}
    \STATE SC.add(V.dot)
    \STATE atomic-write([SC, TS, V])
\ELSE [SC seen V.dot]
	 \STATE atomic-write-if-changed([SC, TS])
\ENDIF
\end{algorithmic}
\end{algorithm}

\section{Appendix: Evaluation} \label{appendix: results}
In-progress evaluations are being run on many setups and environments. For this
paper, we used an Amazon EC2 \textit{cc2.8xlarge}\cite{ec2}, compute-optimized,
64-bit instance type, which includes 2 Intel Xeon processors, each with 8 cores,
3.37 TB of internal storage, and 60.5 GB of RAM. For our
\textit{Riak Sets} and \textit{Deltas}\cite{riakdtdelta} runs,
we used a 4-node Riak/Riak-core cluster built atop on Basho's
\textit{OTP\_R16B02\_basho8} fork\cite{fork} of Erlang OTP.

To exercise our clusters, we used Basho Bench\cite{basho_bench}, a benchmarking
tool created to conduct accurate and repeatable stress tests and produce
performance graphs. The information in table \ref{table1} centers on writes on a
single key with a single worker process. For table \ref{table2}, we used
25 concurrent workers.

The focus of our runs are on write-performance across increasing set
cardinality. We've also included some preliminary results with 1000-key
pareto-distributed\cite{pareto}, 5-minute, read loads, as well as a mixed
write/read load, comparing current Riak Sets against the ongoing Bigset work.

\setlength{\abovecaptionskip}{0ex}
\setlength{\belowcaptionskip}{-4.2ex}

\subsection{Write Runs}
\begin{table}[H]
\centering
\resizebox{\columnwidth}{!}{%
\begin{tabular}{lSSSSSSSSS}
\toprule
\multirow{3}{*}{\textbf{Write Runs}} &
\multicolumn{3}{c}{\textbf{\textit{Riak Sets Avgs.}}} &
\multicolumn{3}{c}{\textbf{\textit{Deltas Avgs.}}} &
\multicolumn{3}{c}{\textbf{\textit{Bigsets Avgs.}}} \\
& {TP} & {M} & {95th} & {TP} & {M} & {95th} & {TP} & {M} & {95th} \\
\midrule
  5k/4bytes & 81.57 & 14.54 & 16.4 & 48.81 & 24.48 & 27.26 & 7094.66 & 7.07 &
  18.71\\
  10k/4bytes & 45.58 & 27.62 & 30.98 & 26.79 & 47.28 & 52.43 & 1040.00 & 0.94
  & 1.12\\
  45k/4bytes & 6.89 & 173.60 & 183.49 & 12.61 & 91.22 & 98.91 & 1060.67 & 0.94
  & 1.11\\
\bottomrule
\end{tabular}
}
\captionsetup{format=hang}
\caption {
  \textbf{Writes} \textit{Cardinality/Size-Per-Element} \newline
  \textbf{TP} - \textit{Throughput} in operations per second (Ops/sec) \newline
  \textbf{M} - \textit{Mean latency} in microseconds \newline
  \textbf{95th} - \textit{95th percentile latency} in microseconds
}
\label{table1}
\end{table}

\begin{figure}[H]
\centering
\resizebox{\columnwidth}{!}{%
\includegraphics[width=20cm, height=14cm]{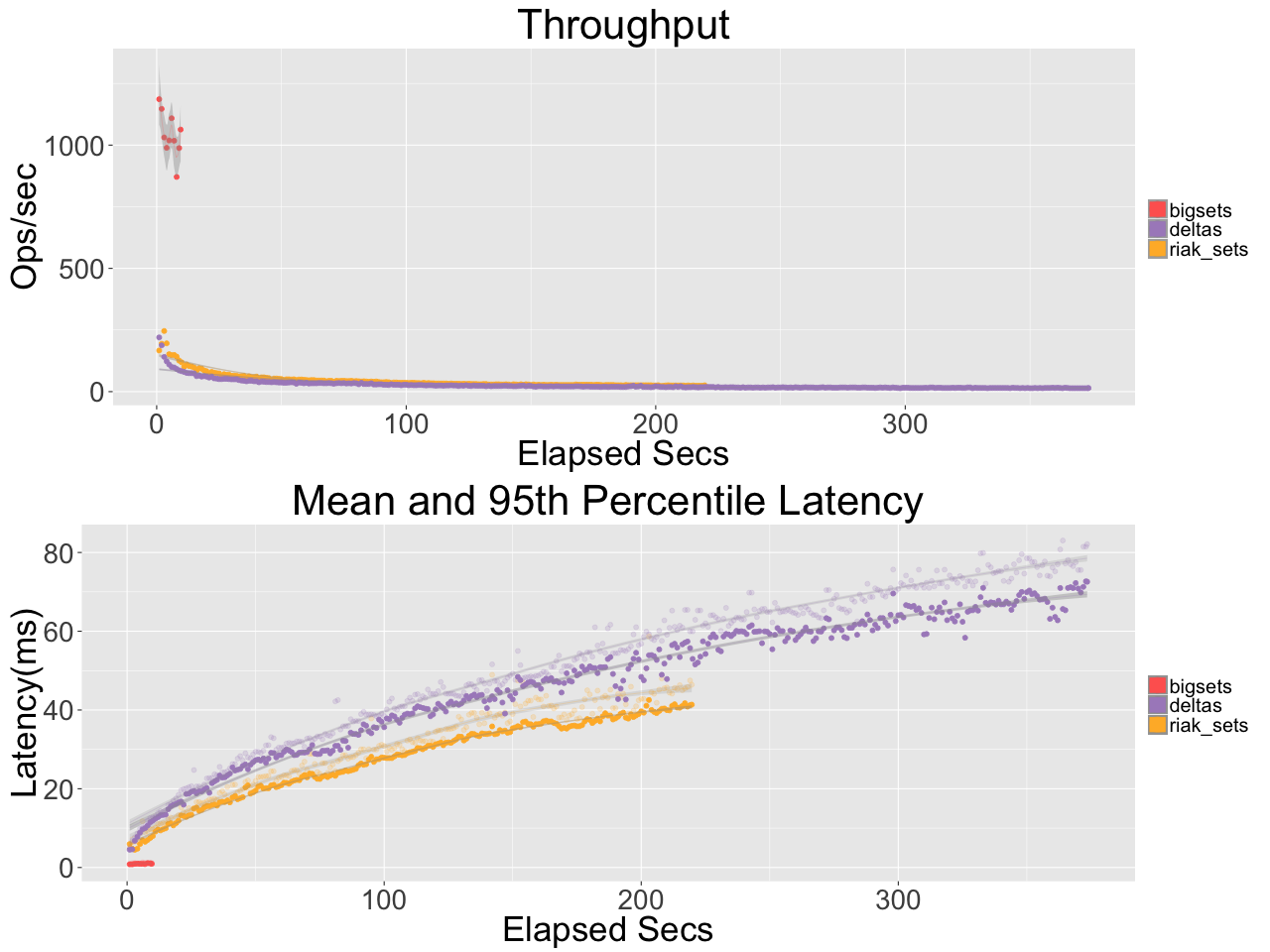}
}
\captionsetup{format=hang}
\caption{Writes on a 10,000-cardinality set of 4-byte elements. \newline
  Bigsets completes getting to the chosen cardinality much more quickly
  than the others.}
\label{figure10k1key}
\end{figure}

\begin{figure}[H]
\centering
\resizebox{\columnwidth}{!}{%
\includegraphics[width=20cm, height=14cm]{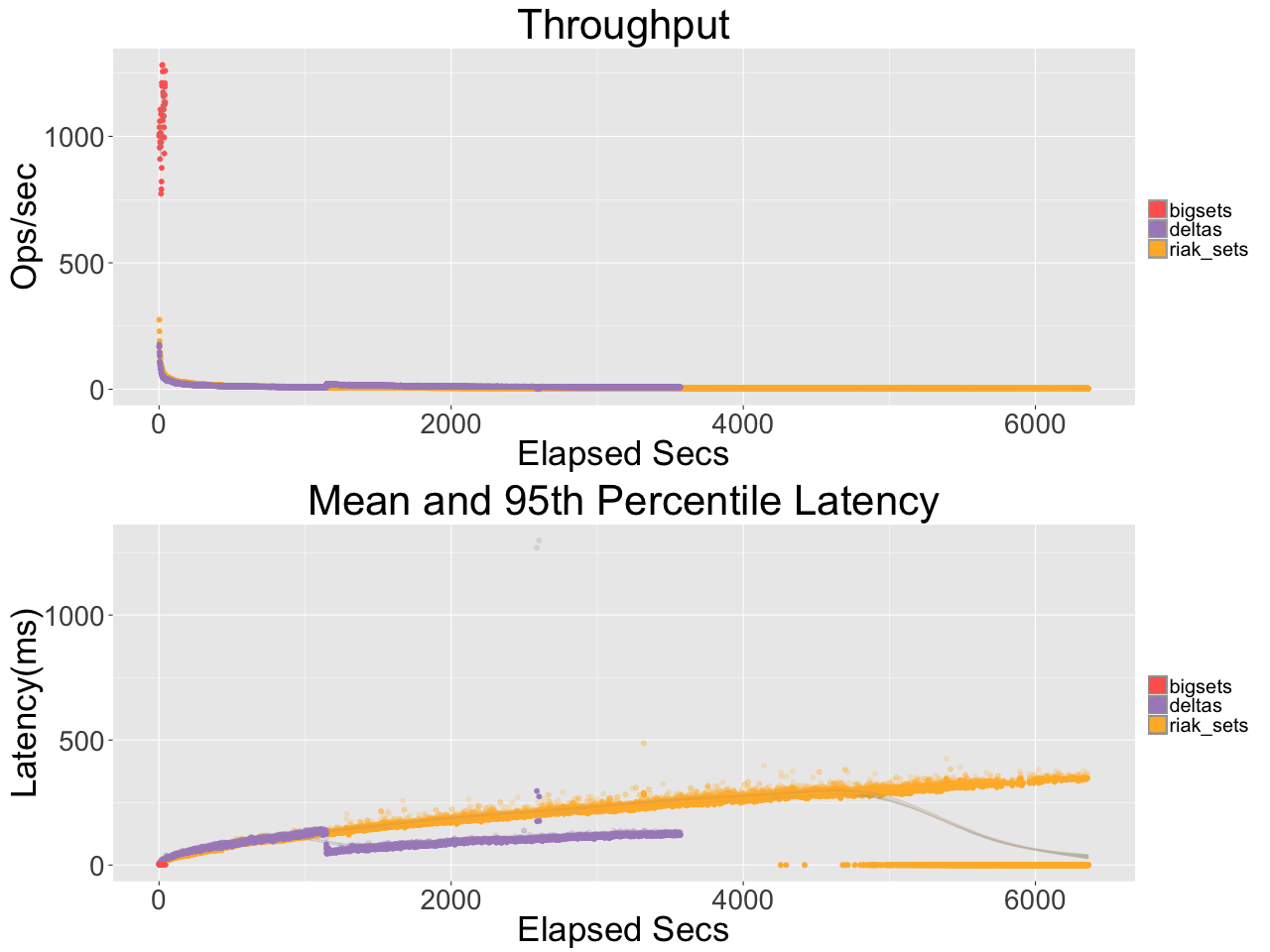}
}
\captionsetup{format=hang}
\caption{Writes on a 45,000-cardinality set of 4-byte elements. \newline
  Like figure \ref{figure10k1key}, it reaches its chosen cardinality much faster.}
\label{figure45k1key}
\end{figure}

\newpage
\subsection{Read and Mixed Runs}
\begin{table}[H]
\centering
\resizebox{\columnwidth}{!}{%
\begin{tabular}{lSSSSSS}
\toprule
\multirow{3}{*}{\textbf{Reads Runs}} &
\multicolumn{3}{c}{\textbf{\textit{Riak Sets Avgs.}}} &
\multicolumn{3}{c}{\textbf{\textit{Bigets Avgs.}}} \\
& {TP} & {M} & {99th} & {TP} & {M} & {99th} \\
\midrule
  1k/1k/4bytes & 6353.83 & 3.93 & 6.85 & 785.23 & 32.70 & 202.18\\
  1k/10k/4bytes & 1266.48 & 19.78 & 32.72 & 221.14 & 116.77 & 632.67\\
  1k/100k/4bytes & 64.11 & 390.96 & 693.71 & 38.55 & 652.56 & 965.75\\
  $\sim$1k/Mix-\textbf{R} & 1198.45 & 4.36 & 40.60 & 2062.72 & 16.48 & 121.97\\
  $\sim$1k/Mix-\textbf{W} & \textquotedbl & 40.00 & 1301.82 & \textquotedbl &
  14.40 & 202.50\\
\bottomrule
\end{tabular}
}
\captionsetup{format=hang}
\caption {
  \textbf{Reads/Mixed} \textit{Keys/Cardinality/Size-Per-Element} \newline
  \textbf{Mix} - 60/40 write-to-read ratio for 5 minute \textit{Write/Read}
  \newline
  \textbf{TP} - \textit{Throughput} in operations per second (Ops/sec) \newline
  \textbf{M} - \textit{Mean latency} in microseconds \newline
  \textbf{99th} - \textit{99th percentile latency} in microseconds
}
\label{table2}
\end{table}

\begin{figure}[H]
\centering
\resizebox{\columnwidth}{!}{%
\includegraphics[width=20cm, height=14cm]{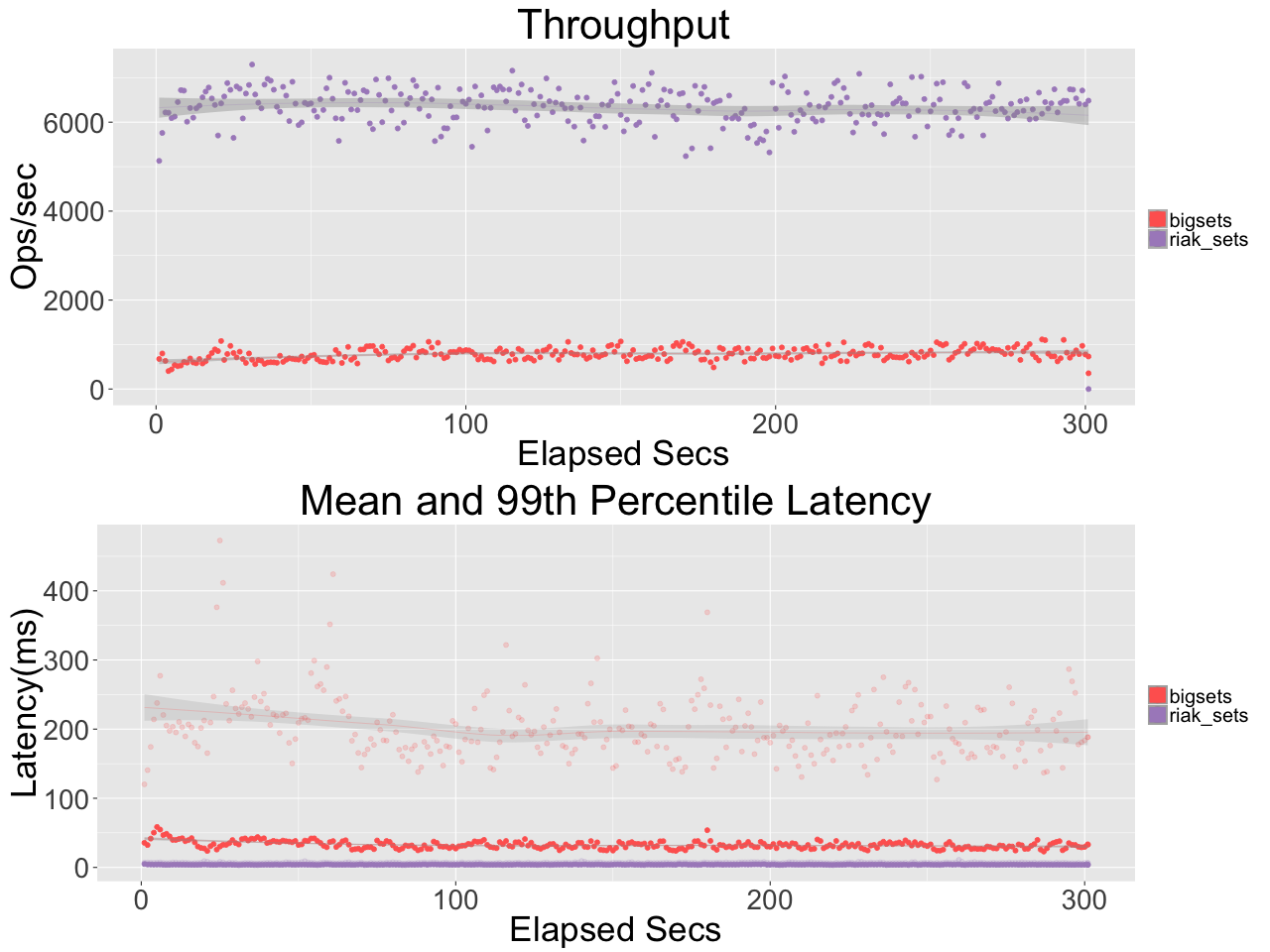}
}
\caption{Reads on a 1000-cardinality set of 4-byte elements.}
\label{figure1kreads}
\end{figure}

\begin{figure}[H]
\centering
\resizebox{\columnwidth}{!}{%
\includegraphics[width=20cm, height=14cm]{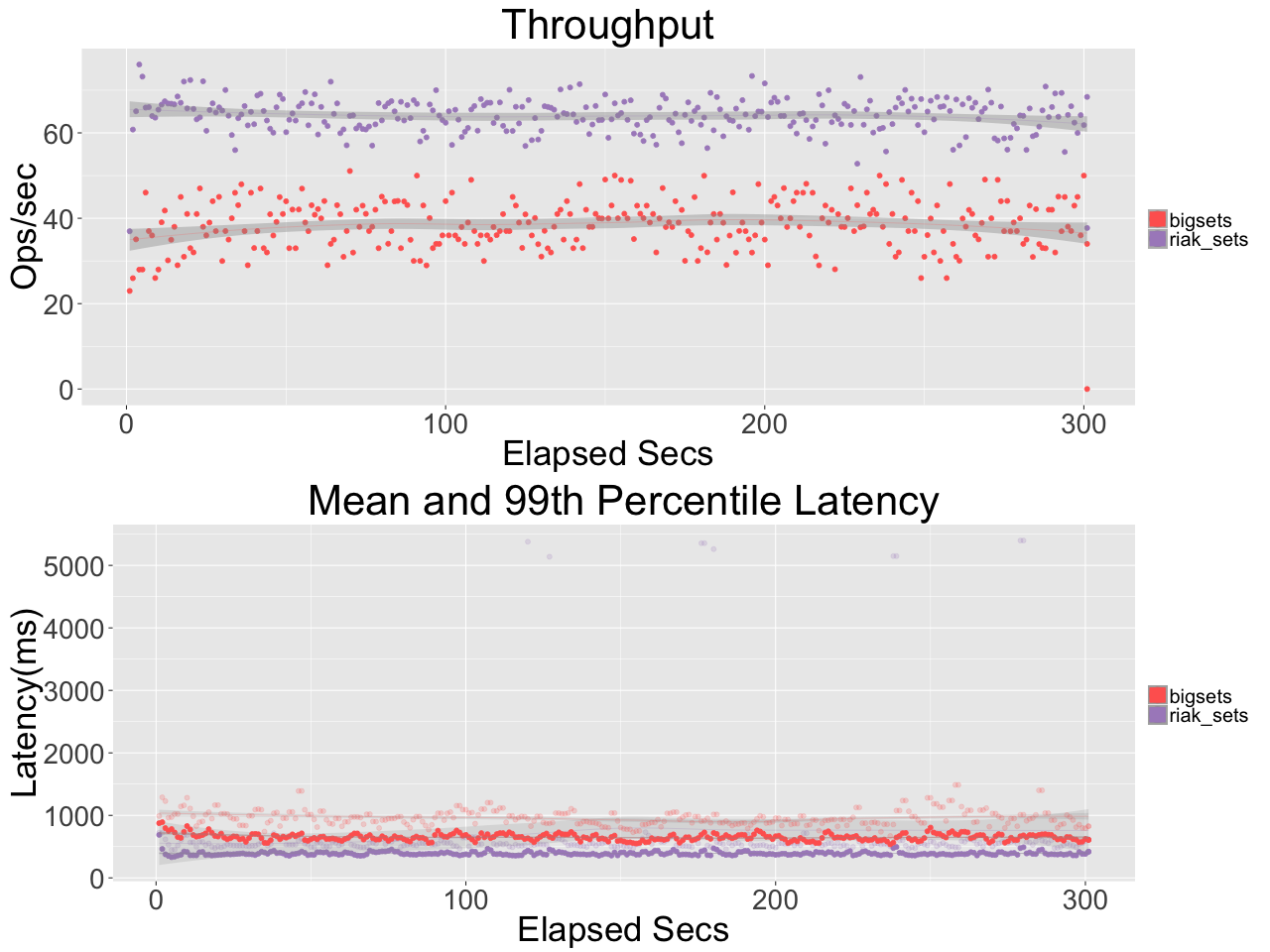}
}
\caption{Reads on a 100,000-cardinality set of 4-byte elements.}
\label{figure100kreads}
\end{figure}

\begin{figure}[H]
\centering
\resizebox{\columnwidth}{!}{%
\includegraphics[width=20cm, height=14cm]{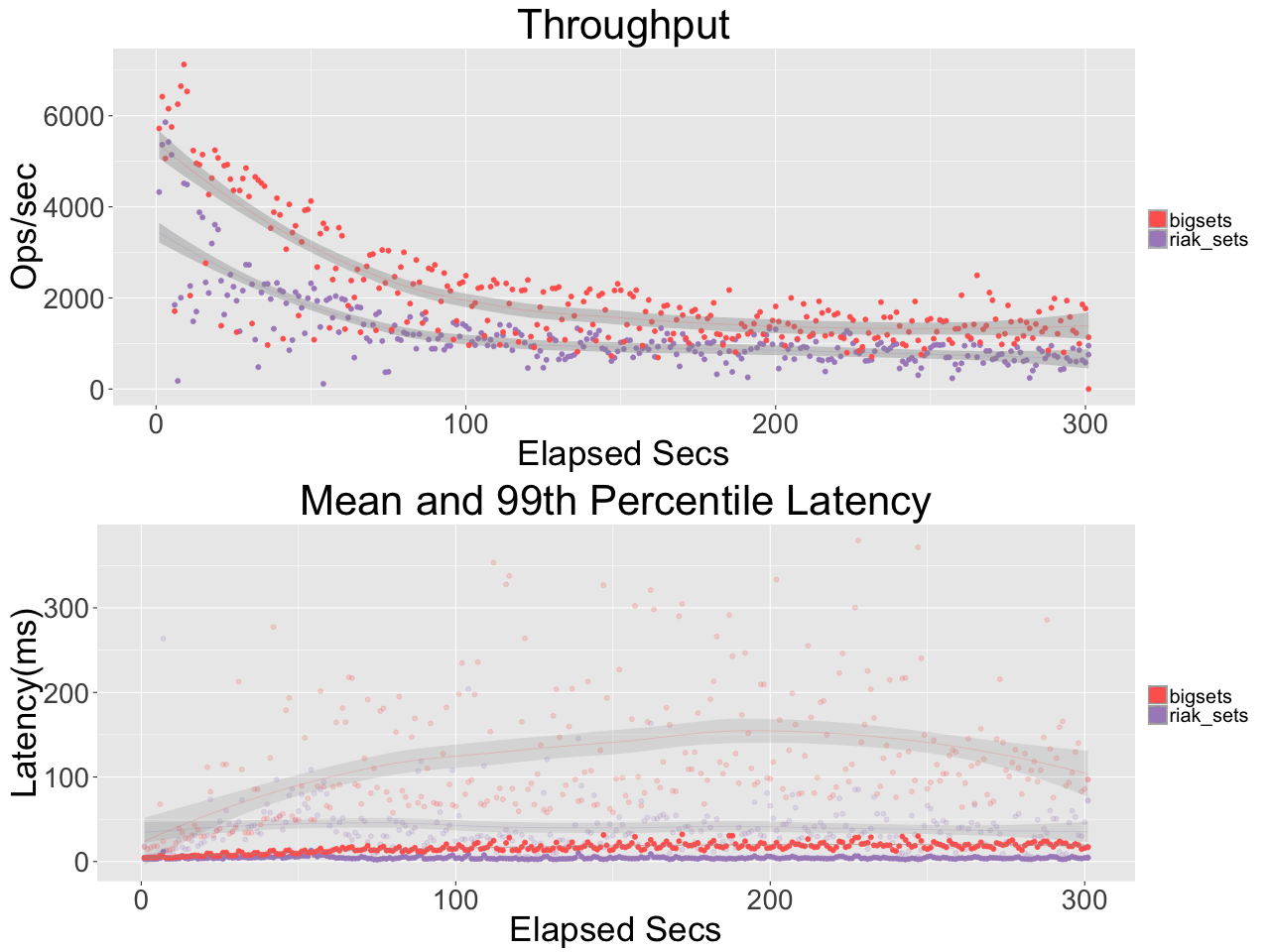}
}
\caption{60/40 write-to-read ratio with read-latencies below.}
\label{figuremixed}
\end{figure}

\end{document}